\documentclass{iaus}
\usepackage{graphicx}

\title[The Chemistry of the Local Group]
{The Chemistry of the Local Group}

\author[Gibson]
{Brad K. Gibson}

\affiliation{Centre for Astrophysics, University of Central Lancashire,
Preston, PR1~2HE, UK \break email: bkgibson@uclan.ac.uk}

\pubyear{2007}
\volume{241}
\pagerange{1--4}
\date{?? and in revised form ??}
\jname{Stellar Populations as Building Blocks of Galaxies}
\editors{A. Vazdekis \& R. Peletier, eds.}
\begin{document}

\maketitle

\begin{abstract}
Simulations of the chemical enrichment histories of ten Local Group (LG) dwarf 
galaxies are presented, employing empirically-derived
star formation histories (SFHs), a rich network of isotopic and elemental
nucleosynthetic yields, and a range of prescriptions for supernova 
(SN)-driven outflows.  Our main conclusions are that (i) 
neutron-capture element patterns (particularly that of Ba/Y) suggest a
strong contribution from low- and intermediate-mass stars (LIMS), (ii) neutron
star mergers may play a relatively larger role in the nucleosynthesis
of dwarfs, (iii) SN feedback alone can explain the observed gas fraction
in dwarf irregulars (dIrrs), 
but dwarf spheroidals (dSphs) require almost all their
gas to be removed via ram pressure and/or tidal stripping, (iv) the
predicted heavy Mg isotope enhancements in the interstellar medium (ISM)
of dwarfs may provide an alternate solution to claims of a 
varying fine structure (v) the gas lost from dwarfs
have O,Si/C abundances in broad agreement with intergalactic medium
abundances at redshifts 2$<$$z$$<$4, and (vi) the chemical properties
of dSphs are well-matched by preventing galactic winds from re-accreting,
whilst those of dIrrs are better-matched by incorporating 
metallicity-dependent cooling and re-accretion of hot winds.  Finally,
doubts are cast upon a claimed association between LG
dSph UMaII and HVC Complex~A.
\end{abstract}

\firstsection % if your document starts with a section,
\section{Introduction}

Due to their low mass, dwarf galaxies are unique laboratories for 
studying the interplay between violent feedback (eg.
SN-driven outflows and starbursts), relatively 
quiescent inflow of both primordial and re-accreted, radiatively-cooled,
metal-enhanced wind material, and the hierarchical assembly of these 
building blocks into more massive systems.  Dwarfs may also give rise
to distant high-column density gas clouds seen in absorption towards even
more distant QSOs.  Further, LG dwarfs are one of the 
few places in the Universe, outside the Milky Way, where (i) individual 
stellar abundances can be derived and (ii) empirical
SFHs can be derived from the colour-magnitude diagrams of resolved
stellar populations.  Environment clearly plays an important role in 
shaping the SFHs and stellar populations of dwarfs, with gas-poor, 
predominantly old, dSphs clustered near the Milky Way, and gas-rich
dIrrs, with more extended SFHs, more widely distributed.

To investigate the differences between dSphs and dIrrs, we have simulated 
the chemical evolution of ten LG dwarfs, exploring a wide range
of prescriptions for SN feedback, cooling and re-accretion of galactic
winds, and nucleosynthetic yields (including neutron capture elements). 
A unique aspect of our efforts is the (attempted) elimination of both
star formation and infall as "free parameters".  We end (Appendix~A)
with a cautionary
note regarding the claimed association between one of the newly-discovered
LG dwarfs - Ursa~Major~II (UMaII) - and 
High-Velocity Cloud (HVC) Complex~A.

\section{Dwarf galaxy models}

Chemical enrichment histories 
were simulated using an updated version of {\tt GEtool} (Fenner \& Gibson
2003); enhancements included: (i)
SNe-driven outflows, (ii) empirically-derived SFHs (Dolphin et~al. 2005), 
(iii) semi-empirically-constrained infall rates,\footnote{In the sense
that the infall rate was "tuned" to ensure the gas density and
SFHs adhered to the Kennicutt (1998) law.}, including
a provision for the incorporation of diluted,
metallicity-dependent, radiatively-cooled, re-accreted
wind material, and (iv)
inclusion of both rapid- and slow-neutron capture nucleosynthetic 
yields.\footnote{The s-process yields were derived using the Torino Group's
post-processing methodology, applied to the Monash Group's yields,
as described by Fenner et~al. (2006).}  We have relaxed the assumption 
that differential galactic winds act in proportion to the current star
formation rate (eg. Lanfranchi et~al. 2006), instead linking
directly to the underlying SNe rates.
We also allow for the mass-loading of cold, entrained, 
ISM gas to the SNe-driven galactic outflows, a key parameter
determining the metallicity of the outflows.

\section{Results}

\subsection{Stellar Component}

Fig.~1 summarises the observed and predicted abundance ratios for 
17 elements for the current sample of ten LG dwarfs.  The loci
of small dots in each panel correspond to simulated model stars (for our
template grid of models) - 
for the sake of brevity, loci of both dSph and dIrr
models have been overlaid.\footnote{Roughly speaking, 
the loci overlie one another, with the separation only becoming 
apparent at [Fe/H]$>$$-$1, where the dots are essentially associated only
with simulated dIrr stars.}  The larger symbols in each panel 
correspond to the extant data for each galaxy listed in the upper left
panel.

Broadly speaking, the template models reproduce many of the features
common to the dwarfs, including the (i) low absolute metallicities, 
(ii) relatively low ratios of [$\alpha$,Sc,Mn, Cu,Y/Fe], and (iii) trends
of [Ba/Eu,Y], indicative of significant pollution from LIMS.
The latter is understood in our models as 
a consequence of the strongly metallicity-dependent s-process
asymptotic giant branch (AGB) yields.  At low metallicities, the ratio of
the neutron flux (provided by the $^{13}$C pocket) to seed nuclei
is high, and the s-process can operate efficiently to produce a 
bottleneck at Pb.  With increasing metallicity, there are fewer
neutrons per seed nuclei and the s-process tends to produce relatively
fewer heavy s-elements and more lighter s-elements.  This should lead to a 
steep rise in Ba/Y with increasing Fe/H in systems with a significant
AGB contribution.

General failures of the model are self-evident and reflect
our poor understanding of certain stellar evolutionary nucleosynthetic
pathways (note: no \it a posteriori \rm scaling of the chemical evolution 
models or yields has been applied).  These failures have been
discussed by Fenner et~al. (2006) and include the (i) 
significant overproduction of Na and Ni, (ii) mild underproduction
of Mg, and (iii) inability to reach 
[Eu/Fe]$\sim$$+$1.\footnote{Alleviated by allowing a modest r-process
contribution from neutron star-neutron star and neutron star-black hole
binaries, as we will show in a future study; the models shown here only 
associate the r-process with SNe~II.}

\begin{figure}
 \includegraphics[scale=0.54]{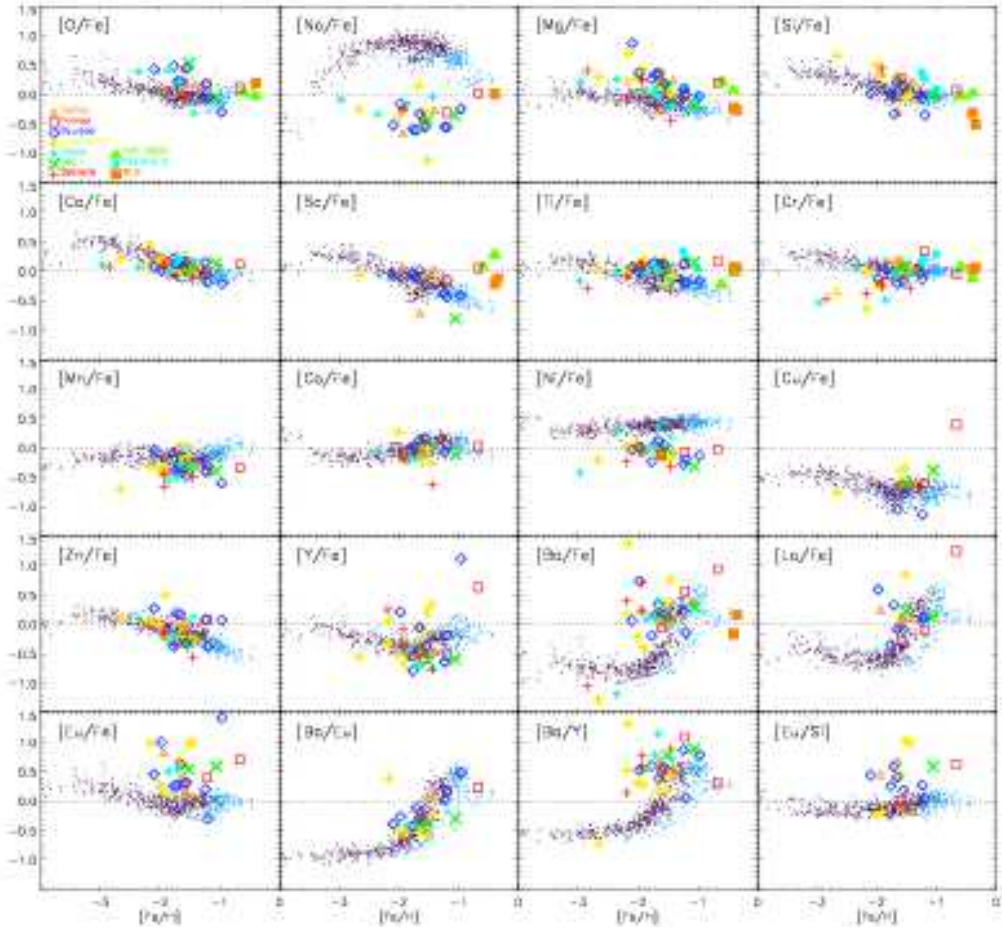}
  \caption{Predicted abundances for the ten template models, 
	incorporating SN feedback and galactic winds (small circles), 
	plotted against stellar observations (larger symbols).  A 
	random uncertainty of $\pm$0.1~dex was imposed upon model
	star abundances to mimic observational errors.}
\end{figure}

\subsection{The Connection to the IGM}

Using the template models and their predicted outflow properties, 
we determined the redshift evolution of 
[O,Si/C] in the intergalactic medium (IGM), in order to compare with that
observed in the IGM at redshifts 2$<$$z$$<$4 
([O,Si/C]$\approx$$+$0.4$\rightarrow$$+$0.7).  The expected pollution from
LG dwarfs was evaluated by summing the mass of metals ejected from 
our grid of dSph and dIrr models and applying a weighting factor to 
account for their observed frequency in the LG.  The ratios 
thusly predicted over this redshift range match the observations regardless of
stripping assumptions, and are fairly robust to model parameters and
the relative number of dIrrs to dSphs.

\subsection{Implications for a Varying Fine Structure Constant}

It has been suggested that anomalous Mg isotopic abundances within 
redshift $z$$<$2 QSO absorbers could mimic the effect of a varying
fine structure constant $\alpha$.  An initial mass function (IMF) biased 
towards intermediate mass stars early in the evolution of a Milky Way-like
system could produce the requisite factor of ten enhancement needed to
reach such extreme $^{25,26}$Mg/$^{24}$Mg ratios, although at the 
expense of introducing peculiar (and unobserved)
nitrogen abundances and carbon isotope ratios (Fenner et~al. 2005).
Conversely, the neutral gas of our simulated dwarfs galaxies bears the 
imprint of LIMS pollution, provided there has been
preferential loss of SNe ejecta, without the need to invoke a finely-tuned
non-standard IMF.  We find that the ISM of our model dSphs readily 
populate the enhanced $^{25,26}$Mg/$^{24}$Mg regime required, but by 
redshift $z$$<$2, most are expected to have lost so much of their
gas as to fall below the neutral hydrogen (HI) column density 
detection limit of QSO absorber studies.  The better retention of gas
by dIrrs could afford them greater representation in the low redshift sample.
Two of our three dIrr models achieve $^{25,26}$Mg/$^{24}$Mg ratios that
fall within the range that could account for the varying-$\alpha$
measurement.  Having said that, these two models assume that SNe-driven
winds are permanently lost; only a modest fraction of the wind needs to be
re-accreted to bring the $^{25,26}$Mg/$^{24}$Mg ratio back down to near-solar.

%\section{Conclusions}

\appendix
\section{The Orphan Stream and HVC Complex~A}

The so-called "Field of Streams" has proven to be a boon
for LG dwarf studies, leading to the discovery of many new dwarfs,
and the fascinating Orphan Stream (Belokurov et~al. 2007).  Belokurov et~al,
Fellhauer et~al. (2007), and Jin \& Lynden-Bell (2007) each note the
spatial and kinematical coincidence of HVC Complex~A with
the predicted orbit of one of these new dwarfs, UMaII.
with Fellhauer et~al. making the 
reasonable suggestion (via the use of N-body simulations)
that the progenitor of UMaII was likely to have had a {\it dynamical} mass of 
$\sim$3$\times$10$^{5}$~M$_\odot$ (and certainly $<$10$^6$~M$_\odot$)
and now lies at a heliocentric 
distance of $\sim$30~kpc (with an absolute magnitude of
$\sim$$-$7.5).  While we find it likewise tempting to associate UMaII
and the Orphan Stream with Complex~A, the latter association appears difficult
to reconcile with the following points:
\begin{itemize}
\item Complex~A lies 4-10~kpc from the Sun, and not 30~kpc
\item assuming UMaII adheres to the luminosity-metallicity relation, 
much like the other LG dSphs, one would expect a 
metallicity $<<$1\% solar, for its luminosity, while Complex~A almost
certainly possesses a metallicity $>$10-20$\times$ that value
\item the $\alpha$/Fe ratio within Complex~A is likely enhanced, as
in Complex~C, and
consistent with a SNeII signature, while present-day LG dSphs
typically present sub-solar $\alpha$/Fe
\item perhaps most difficult to reconcile is the fact that at
a putative distance of 30~kpc, the HI mass of Complex~A
would be $\sim$10$^7$~M$_\odot$ (even neglecting the known and substantial
ionised gas components associated with Complex~A), $\sim$30$\times$ 
the dynamical mass of the putative UMaII progenitor, alone.  Even at 
the lowest allowable distance (4~kpc), the HI 
mass of Complex~A already equates to the putative UMaII progenitor's
dynamical mass,
leaving no room for a dark matter or stellar component.
\item a further $\sim$10$^6$~M$_\odot$ of stars would also be required
to enrich the 10$^7$~M$_\odot$ of gas to 10\% solar metallicity,
only exacerbating this mass "problem"
\end{itemize}
\smallskip
In consort, while it is tempting to link Complex~A with UMaII and the 
Orphan Stream, the problem appears more complicated than one of simply
recovering spatial and kinematical coincidence!

\begin{acknowledgments}
This work could not have been undertaken without the remarkable
skills of Yeshe Fenner; her contributions are gratefully acknowledged, as
is the ongoing guidance of Leticia Carigi, 
Andrea Marcolini, St\'ephanie Courty, and Patricia S\'anchez-Bl\'azquez.
\end{acknowledgments}

\end{document}